\documentclass[%
 reprint,
 amsmath,amssymb,
 aps,
]{revtex4-2}

\usepackage{graphicx}
\usepackage{dcolumn}
\usepackage{bm}
\usepackage{xcolor}
\usepackage{float}
\usepackage{hyperref}

\begin{document}


\title{Stacking-Mediated Diffusion of Ruthenium Nanoclusters in Graphite}

\author{James G. McHugh}
\email{james.mchugh@manchester.ac.uk}
\affiliation{Dept. of Physics \& Astronomy, University of Manchester, \\Oxford Road, Manchester M13 9PL, United Kingdom\\}
\affiliation{National Graphene Institute, University of Manchester, Booth St E, Manchester M13 9PL, United Kingdom}

\author{Pavlos Mouratidis}
\email{p.mouratidis@lboro.ac.uk}
\affiliation{Dept. of Chemistry, Loughborough University\\
 Epinal Way, Loughborough LE11 3TU, United Kingdom}

\author{Kenny Jolley}
\email{k.jolley@lboro.ac.uk}
\affiliation{Dept. of Chemistry, Loughborough University\\
 Epinal Way, Loughborough LE11 3TU, United Kingdom}

\date{\today}

\begin{abstract}
The diffusion, penetration and intercalation of metallic atomic dopants is an important question for various graphite applications in engineering and nanotechnology. We have performed systematic first-principles calculations of the behaviour of ruthenium nanoclusters on a graphene monolayer and intercalated into a bilayer. Our computational results show that at a sufficiently high density of single Ru atom interstitials, intercalated atoms can shear the surrounding lattice to an AA stacking configuration, an effect which weakens with increasing cluster size. Moreover, the interlayer stacking configuration, in turn, has a significant effect on cluster diffusion. We therefore find different trends in diffusivity as a function of cluster size and interlayer stacking. For monolayer graphene and an AA graphene bilayer, the formation of small clusters generally lowers diffusion barriers, while the opposite behaviour is found for the preferred AB stacking configuration. These results demonstrate that conditions of local impurity concentration and interlayer disregistry are able to regulate the diffusivity of metallic impurities in graphite.
\end{abstract}

\maketitle


\section{Introduction}

\par The energetic and dynamical properties of transition metal impurities adsorbed on top of graphene and intercalated between the layers of graphite is a recurring topic of considerable interest in materials science. Foreign elemental impurities are one of the most promising ways to modify the physical properties of pristine graphene \cite{brey2015}, and they hold considerable promise in engineering desirable electronic phases \cite{hu2012}. For example, transition-metal doping of monolayer graphene (MLG) and bilayer graphene (BLG) can be exploited to increase the weak intrinsic spin-orbit effects of the native carbon atoms \cite{marchenko2012, frank2017, slaw2019}, allowing the engineering of novel quantum states with advantageous transport properties and prospective applications as topological insulators or in quantum computing \cite{tian2017, lee2021, gilbert2021}.

\par Intercalated metallic species are also of interest in their own right, and the layered structure of van der Waals materials, such as graphite, provides an excellent platform to grow quasi-two dimensional sheets of selected transition metals \cite{yakub2019,huang2020,rosales2020,Han2017}. These two-dimensional transition metal sheets have many desirable properties, which are greatly enhanced by their lower-dimensional topology. Due to the higher (2D) bulk to surface ratio and the associated change in coordination, layered metals can completely change electronic properties such as band gaps and transport properties \cite{yang2015, nevalaita2018}. Two-dimensional metallic layers have been grown underneath the top monolayer of highly-oriented pyrolytic graphite (HOPG) \cite{LiiRosales2018, rosales2020}, a process which is known to proceed via the diffusion of impurities through lattice defect "entry portals" (monovacancy and multi-vacancy complexes) \cite{LiiRosales2020, Han2017}, and vacancy sites are also known to promote the intercalation of other elemental species such as Cs and Dy \cite{Zhou2018, Bttner2011}.

\par Transition metals are also important from the perspective of graphite applications. Some of the most important nuclear fission products are transition metals \cite{wojc2020}, and the penetration of these impurities into the bulk and subsequent diffusion through the graphite lattice is a pressing problem in the design of new reactors and in the assessment of safety and decommissioning of retired reactors \cite{Marsden2017,Wen2008, Marsden2016}. Nuclear graphite occurs in various distinct grades, which are defined both by anisotropy and porosity, both of which are proportional to the size and density of graphite crystallites \cite{Wen2008, Marsden2016, Kane2011, Zhou2017}. It is likely that penetration into and diffusion within these grades proceeds via a similar entry portal mechanism, as there are ample defect sites which can serve as entry sites to aid the diffusion of intercalated species \cite{McHugh2022}.

\par In this context, ruthenium is a particularly interesting atomic impurity in graphite. In isotropic nuclear graphite grades, diffusivity experiments demonstrate an unusual turnaround, where above a critical temperature of approximately 700$^{\circ}$C the diffusivity temporarily decreases, which is counter to the anticipated trend of increasing diffusivity with increasing temperature \cite{Graydon}. The intercalation and embedding of ruthenium into the bulk and top layer of graphite also displays a sharp temperature dependence \cite{LiiRosales2018}. Here, there are distinct morphologies depending on the annealing temperature. At lower temperatures, atoms are evenly distributed above the capping layer, while above a critical temperature, which is again around 700$^{\circ}$C, isolated ruthenium atoms penetrate and agglomerate into more extensive, large and immobile clusters. These encapsulated clusters range in diameter and height and form an identical Moir\'{e} pattern to graphene deposited on the Ru (100) surface system \cite{LiiRosales2018}.

\par Central to understanding these phenomena is the clustering of Ru atoms on graphene and graphite, and the subsequent effect this has on diffusion. While there have been a variety of studies of the energetics of single adsorbed and intercalated elemental species, the effect of clustering on the dynamics of transition metal diffusion has not been considered in any significant depth to the authors knowledge. However, these factors are liable to play a very significant role in realistic diffusion processes, and properly accounting for dynamics is essential in assessing actual physical behaviour. Furthermore, systematic studies of small metallic clusters can provide insight into the early stages of agglomeration and clustering in graphite. 

\begin{figure*}[htbp]
\includegraphics[width=0.85 \textwidth, height=\textheight,keepaspectratio, trim = 0cm 0cm 0cm 0cm, clip = true]{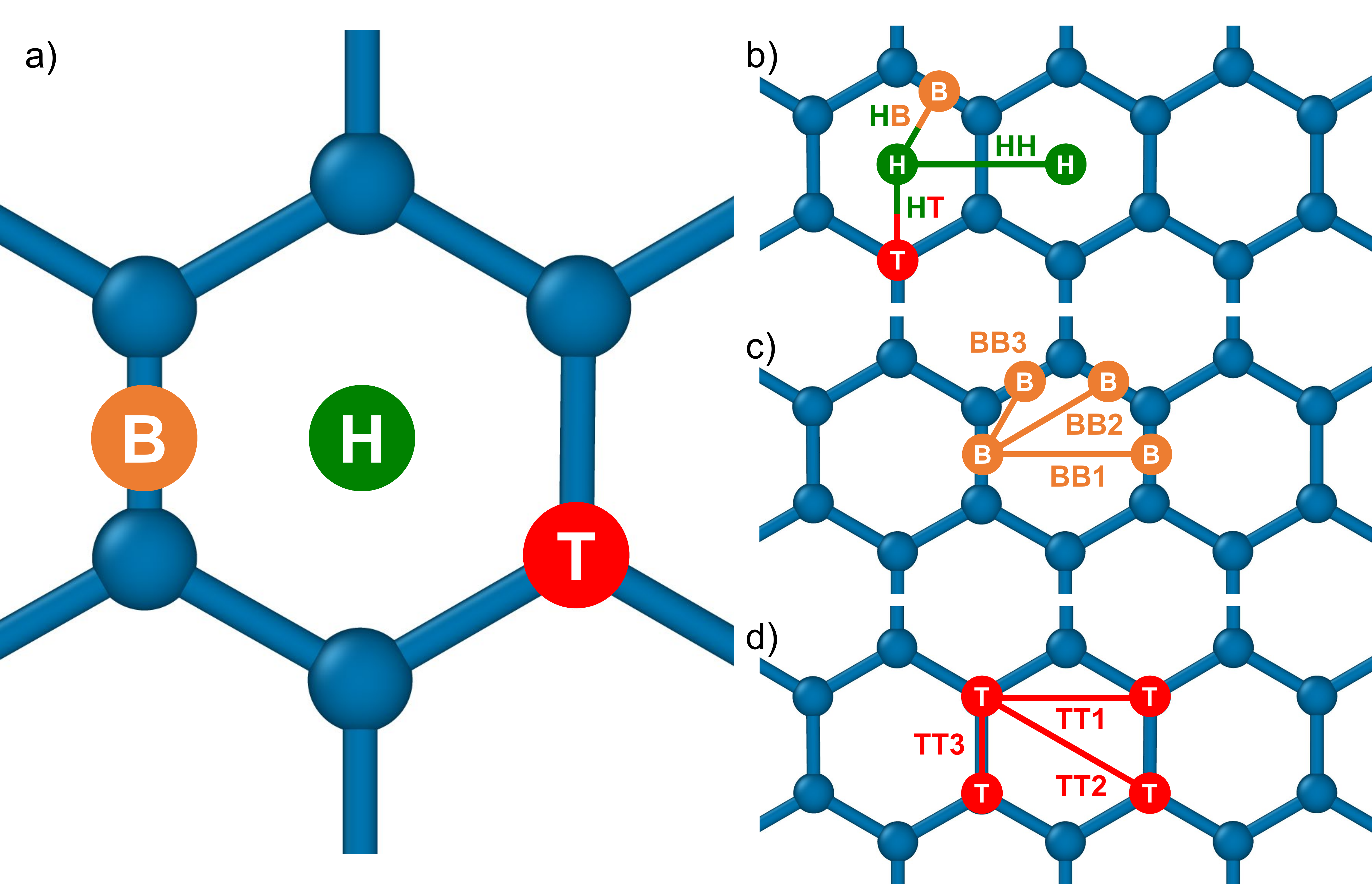}
\caption{High symmetry Ru atom and dimer positions on graphene. (a) Top (T), bond (B) and hollow (H) sites of an adatom adsorbed on graphene. (b)-(d) Various combinations of the single atom positions are used to fully investigate the morphology of adsorbed Ru dimers.}
\label{fig1}
\end{figure*}

\par In this work, we present a systematic study of the interaction of Ru atom nanoclusters ($n = 1,2,3,4,5,7$) on MLG, BLG and bulk graphite, using accurate first-principles calculations to systematically investigate the effect that clustering and local carbon environment have on Ru diffusion in graphite. Ruthenium is chosen both as a representative element and in order to provide new insight into the relevant microstrutural processes underlying transition metal diffusion on the graphite basal (0001) surface, and the penetration and diffusion into the bulk of more complex graphite grades.


\section{Computational Methods}
All DFT calculations have been performed using the Quantum ESPRESSO \textit{ab-initio} package \cite{QE1, QE2, QE3}. The generalized gradient approximation (GGA) was used to approximate the exchange-correlation functional, as parameterized by Perdew, Burke, and Ernzerholf (PBE), and Vanderbilt ultrasoft pseduopotentials are employed to approximate the effect of core electrons \cite{Garrity2014}. The wavefunction basis was expanded as a series of plane-waves with maximum cut-off of $E_{cut} = 40$ Ry (544 eV), and a charge density cut-off of $E_{\rho} = 500$ Ry (6803 eV), which are found to be in close agreement to the more highly-converged calculations. A non-zero electron temperature of $kT_B = 0.02$ eV is applied to aid convergence, using a Fermi-Dirac smearing function, and the Brillouin zone is sampled using a Monkhorst-Pack $5 \times 5 \times 1$ \textit{k}-point grid. More highly converged parameters, using a denser $13 \times 13 \times 1$ \textit{k}-point grid and an energy cutoff of $E_{cut}$ = 60 Ry have been applied for density of states calculations.

The Grimme DFT-D2 method has been employed to model the interlayer van der Waals interaction \cite{Grimme2006}. In monolayer and bilayer calculations, a vacuum thickness of 20 {\AA} along the z-direction is found to avoid any spurious self-interactions between periodically repeated images. Structural optimisations were performed until the residual force on each atom is less than 0.01 eV/{\AA} and the energy difference between subsequent iterations is less than 0.01 eV.

The adsorption energy per atom (chemical potential) has been calculated as 
\begin{equation}
    \mu(n) = \frac{E_{tot} - E_{Gr}}{n}  - E_{Ru, fcc},
\end{equation}
where $E_{Gr}$ is the energy of the respective perfect graphitic material (monolayer and AB stacked bilayer graphene), $E_{Ru,FCC}$ is the energy of bulk fcc Ru lattice, taken from a well-converged bulk calculation sampled with a $21 \times 21 \times 21$ \textit{k}-point grid, $E_{tot}$ is the total energy of the fully-optimised combined Ru nanocluster/graphite system, and $n$ is the number of adsorbed or intercalated Ru atoms \cite{LiiRosales2020}. To facilitate the comparison of energies between different bilayer stacking  configurations (different lateral offsets between the graphene layers), we have calculated all intercalation energies relative to the AB-stacked minimum. This expression gives the relative energy per Ru adatom of different ruthenium-carbon configurations. It is defined such that energetically-preferred structures have lower formation energy, and configurations with $\mu < 0$ are preferred over bulk Ru. In this way, we can gain full insight into the relative preference for adsorption, intercalation and clustering both on the monolayer graphene surface and intercalated into the bilayer or bulk lattice.

\section{Results}
\subsection{Energy \& Structures}
In order to understand the interaction of ruthenium nanoclusters with graphite, we will initially focus on the adsorption (on a graphene monolayer), and intercalation (within a bilayer) of small ($n \leq 7$) clusters. By evaluating the size dependence of nanocluster structure and diffusivity, we can extract many of the essential physical factors moderating the behaviour of ruthenium in graphite at low concentrations. All initial calculations were performed in a $6 \times 6$ graphene supercell, for both monolayer and bilayers with AA stacking (AA BLG) and AB stacking (AB BLG) configurations of the graphene sheets. In the bilayer case, carbon atoms are restricted to relax out of plane, so as to maintain the initial stacking configuration. Importantly, intercalation calculations in both the isolated bilayer and similar bulk cells produce markedly similar results, and we anticipate that the results discussed here are also directly relevant to bulk graphite (see S.I. section 2).

\textit{Monolayer} --- 
We first compare the adsorption and intercalation of single-atom, dimer and trimer ruthenium ($n=1,2,3$) nanoclusters. $n=1,2$ relaxation was performed for a number of different initial configurations, which are detailed in Fig. 1. The formation energy of all fully relaxed structures was then evaluated according to Eq. 1, and the lowest-energy configurations are taken to be the most thermodynamically stable. Tables of all formation energies corresponding to these initial condition are provided in the supplementary information. 

For $n=1$, we find that a single Ru atom prefers to adsorb at the H-site  of the graphene sheet (Fig 1 (a)). Here, the interaction between the Ru and C atoms is predominantly covalent, such that a single ruthenium adatom is best able to hybridise valence electrons at the most highly-coordinated, symmetrical hollow site \cite{McHugh2020}. 

\begin{figure*}[htbp]
\includegraphics[width=1\textwidth, height=\textheight, keepaspectratio, trim = 0cm 1cm 1cm 1cm, clip = true]{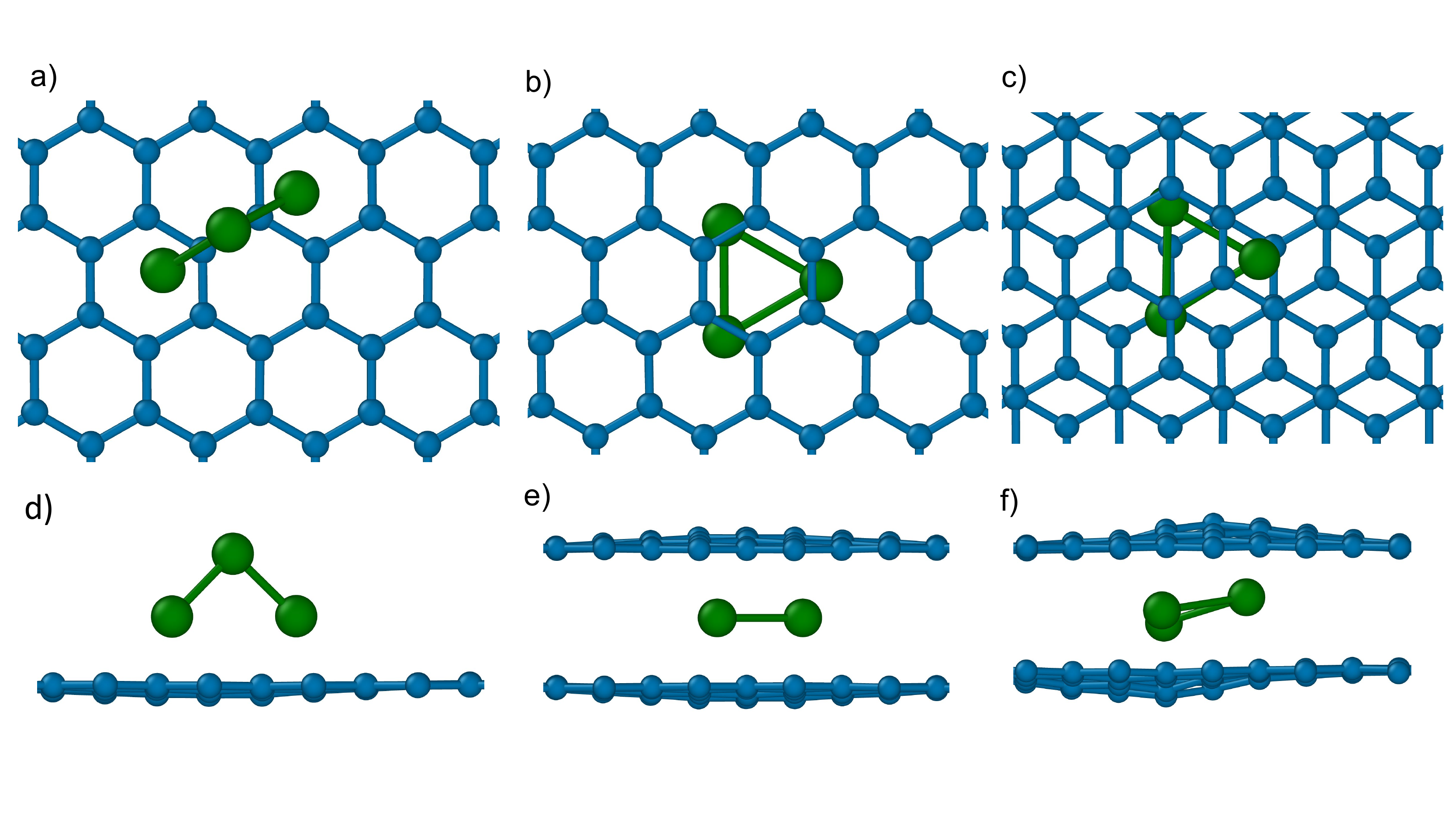}
\caption{Plan and front views of $n=3$ Ru clusters on (a,d) monolayer graphene, (b,e) AA BLG, and (c,f) AB BLG.}
\label{fig2}
\end{figure*}

With increasing Ru cluster size, the constituent ruthenium atoms show significant energetic preference for the formation of metallic bonds with each other, and there is a progressive decrease in the adsorption energy per adatom from 4.82 eV to 4.25 eV, to 3.46 eV for $n=1,2,3$ respectively. This preference is exemplified by the difference between the dimer and trimer structures. As a result of reduced Ru-C hybridization, a pair of Ru adatoms will move away from the hollow site and adopt positions above the C-C bonds, on opposite sides of a graphene hexagon, (configuration BB1 in Fig. 1 (c)), with a slight reduction in Ru-Ru bond length after relaxation. This is similar to the behaviour of an Ir dimer adsorbed on graphene \cite{Barker2019}, which adsorbs in the BB2 configuration due to the relatively shorter Ir dimer bond length. 

For the trimer (Fig 2(a)), there is a more marked departure from the initial structure, and we find that the additional Ru atom is lifted entirely out of the plane, with two Ru adatoms which "carry" the remaining one. They thus effectively form a pair of conjoined dimers, which maximise preferred Ru-Ru interactions while reducing the contact area of the cluster with the graphene sheet.

\textit{Bilayer} --- We now examine the corresponding behaviour of small clusters intercalated within bilayer graphene, and the effect that the stacking has on intercalation energy. In this case, we again find a signficiant tendency to prefer the higher coordination hollow site. This is demonstrated by the preference for a single Ru to sit at the TH-site, which lies above a H-site of one layer and a T-site in the opposite layer, in AB BLG, while For AA BLG, a single Ru atom adopts the site between the two H-sites on both layers. Notably we find a marked preference for AA stackng in our cells, with an intercalation energy of 3.44 eV (3.86 eV) for the AA (AB) bilayer. Hence, at the chosen, relatively high Ru density of one atom per 72 carbon basal plane atoms, the intercalated Ru atoms will shear adjacent sheets and reverse the preferred stacking orientation of perfect graphite, behaviour which is qualitatively similar to graphite intercalation compounds \cite{Dresselhaus2002}.

\begin{figure}[htbp]
\includegraphics[width=0.6\textwidth, height=\textheight, keepaspectratio, trim = 6cm 1cm 2cm 1cm, clip = true]{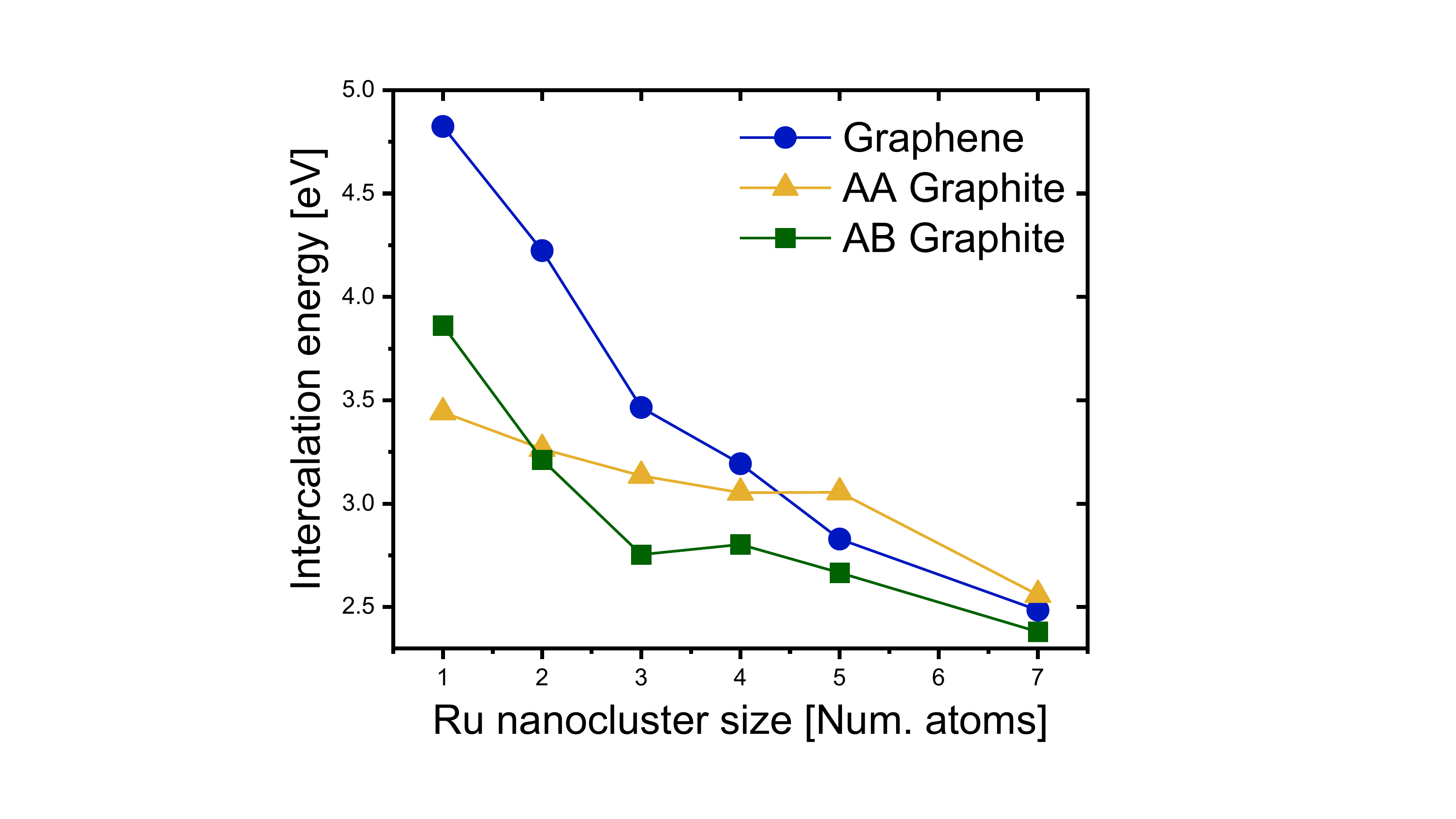} \caption{Adsorption (intercalation) energy per Ru atom of Ru nanoclusters adsorbed (intercalated) on graphene (AA/AB graphite) as a function of Ru nanoclusuter size.} 
\label{fig3}
\end{figure}

\begin{figure}[htbp]
\includegraphics[width=0.45 \textwidth, height=\textheight,keepaspectratio]{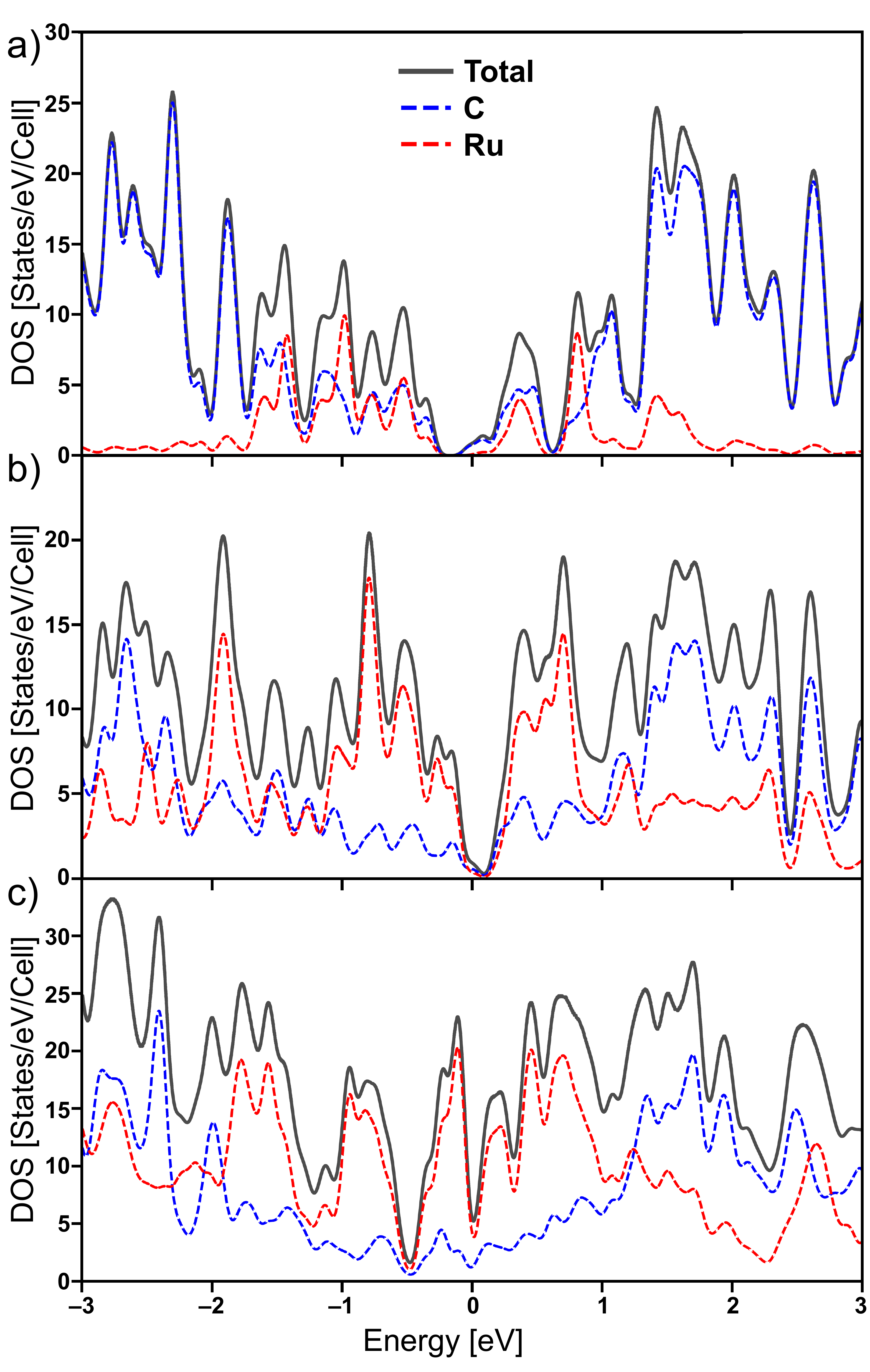}
\caption{Partial density of states of Ru clusters on ML graphene, for (a) $n=1$, (b) $n=4$  and (c) $n=7$.}
\label{fig4}
\end{figure}

Dimer adsorption is relatively similar to the monolayer case, with the Ru-Ru bond lying parallel to the BLG basal plane in both stacking orientations. For AA stacking each dimer atom sits on adjacent H-sites, and there is a large, stretched bond joining them, while for AB stacking they lie on neighbouring TH/HT sites. The two stacking configurations are almost degenerate in energy, and there is a very weak tendency towards AB stacking, with intercalation energies of 3.265 eV and 3.234 eV for the AA and AB bilayers, respectively.

Larger differences between MLG adsorption and BLG intercalation can be seen for the Ru trimer, which is exemplary of the general trends observed for all larger clusters. In this case, it is no longer possible for nanoclusters to relax in the out of plane direction, which is inhibited by the surrounding graphene layers. For both stacking orientations, $n=3$ Ru clusters adopt a flat, triangular shape. In AA BLG, the cluster atoms sit near adjacent occupy symmetrical positions adjacent to the C-C bonds around a graphene hexagon (see Fig. 2(b) and 2(e)), while they inhabit two adjacent TH and one TT site for AB BLG (2(c) and 2(f)). In this case, AB stacking is significantly preferred, with an intercalation energy per Ru atom of 2.75 eV compared to 3.14 eV for the AA bilayer.

\textit{Larger clusters} --- Additional simulations of larger clusters in the same $6 \times 6$ MLG and AA/AB BLG supercells, further exemplify the previously observed trend, as shown in Fig. 3 (Further details are given in S.I. section 1). For cluster sizes of $n \geq 2$, intercalation into the bilayer is preferred for all larger cluster sizes, with a preference for AB stacking. We additionally observe that for $n > 4$, the relative energy difference of adsorbed and intercalated atoms significantly decreases due to the fact that the energy per intercalated Ru atom will tend to 0 eV (i.e., close to that of bulk Ruthenium) with increasing cluster size. For clusters larger than $n=4$,  the thermodynamic driving force for intercalation is therefore substantially reduced in comparison to single atoms, which are able to more substantially reduce their energy by diffusing into the bulk.

\begin{figure}[htbp]
\includegraphics[width=0.5\textwidth, height=\textheight, keepaspectratio, trim = 6cm 0cm 6cm 2cm , clip = true]{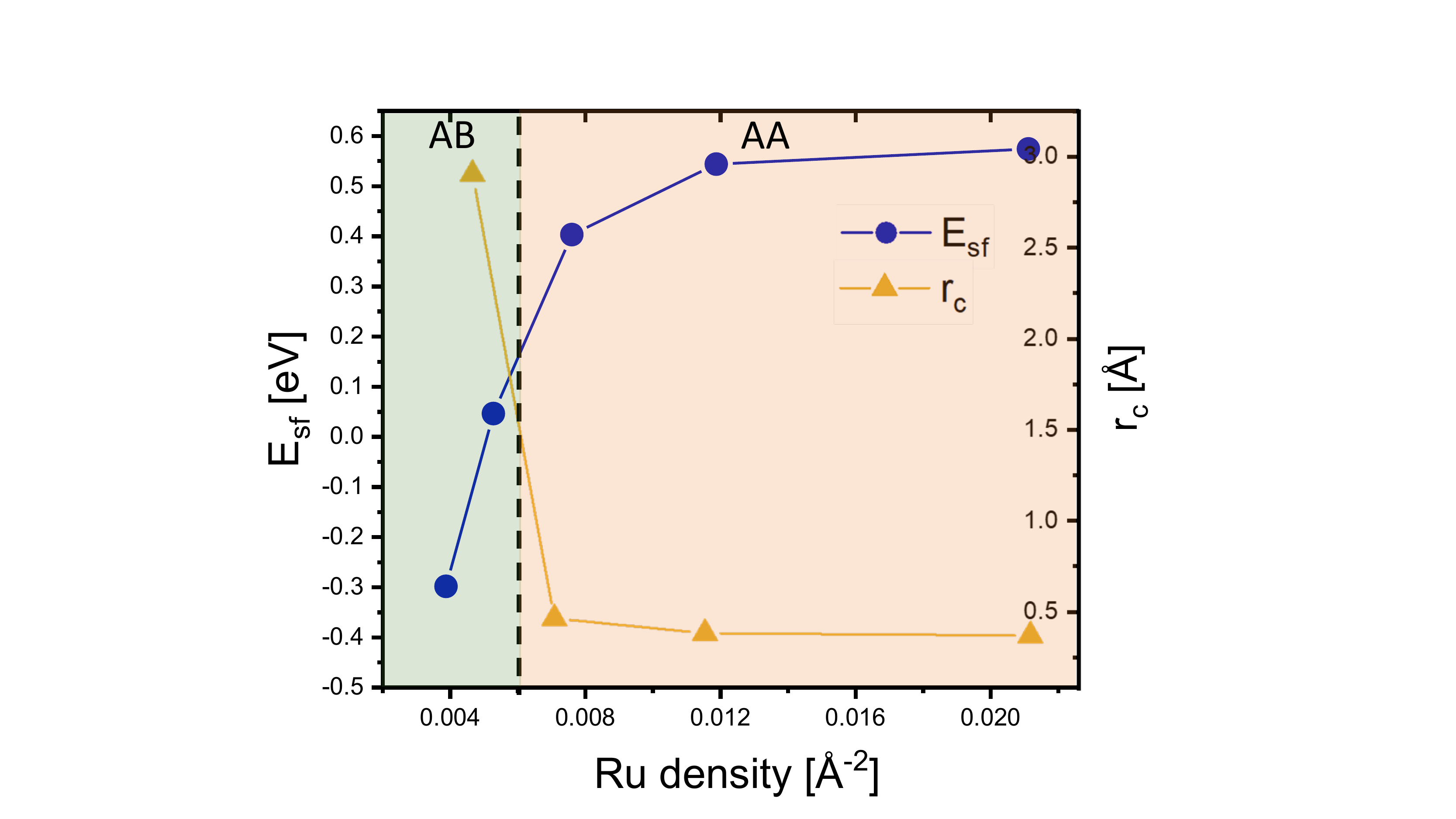}
\caption{Energy difference, $E_{sf} = E_{AA} - E_{AB}$ and critical radius to create a stacking fault, $r_c$, vs ruthenium density for equally-spaced Ru atoms in bilayer graphene.}
\label{fig5}
\end{figure}

\begin{figure*}[htbp]
\includegraphics[width=0.9\textwidth, height=\textheight,keepaspectratio]{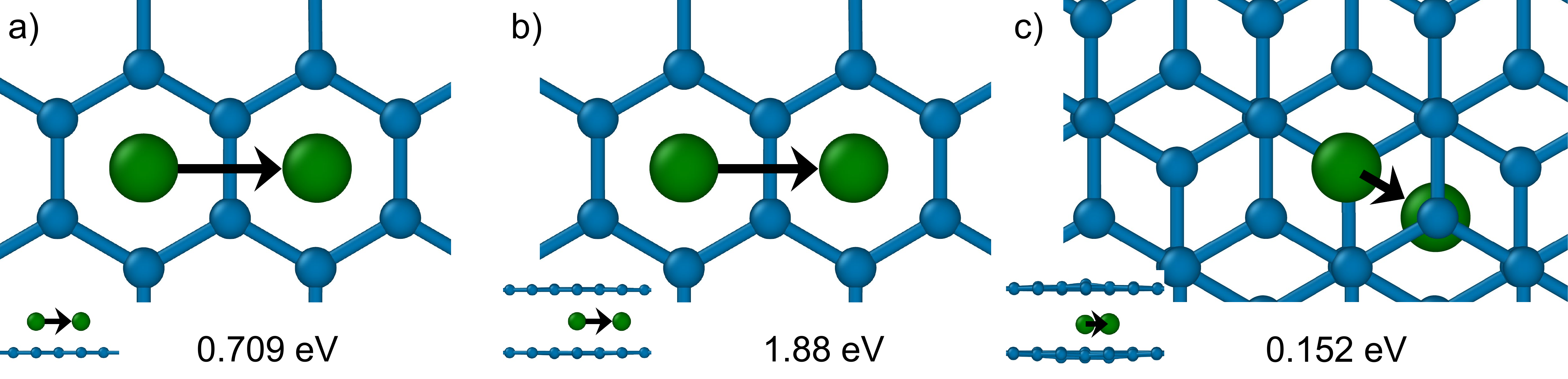}
\caption{Diffusion paths for ruthenium atoms in (a) monolayer graphene (H-site to H-site), (b) AA bilayer graphene (HH-site to HH-site) and (c) AB bilayer graphene (TH-site to TH-site).}
\label{fig6}
\end{figure*}

\textit{Ru density and stacking} ---  The energetic preference for AA-stacking is interesting and deserves further investigation. Trends in adsorption and intercalation can be understood through the partial density of states (PDOS), which allows us to resolve the overlap of electrons associated to the graphitic layers and Ru clusters. This is shown for nanoclusters of size $n=1,4,7$ in Fig. 4, for the monolayer system, which exemplifies the general behaviour. For a single adsorbed Ru, we observe a number of shared peaks in the C and Ru DOS, indicative of shared electrons due to Ru-C covalent bonds. With increasing Ru cluster size this behaviour is less pronounced, with, for example, relatively fewer shared peak for the $n = 7$ cluster due to the preference for Ru-Ru covalent bonding. Interaction between nanoclusters and adjacent graphene layers is then dictated by two opposing factors: interlayer stacking fault energy, which favours AB stacking, and preference of Ru atoms atoms for the H-site. for small Ru clusters ($n=1,2$), the Ru-C interaction weakens pi-bond overlap between graphene sheets, which permits intercalated Ru to overcome the stacking fault energy and pull surrounding graphene sheets to the preferred hollow site. For larger cluster sizes $n \geq 3$, this ability is significantly reduced due increasing Ru-Ru metallic bonding, and the consequent shift away from the preferred H-site.

\textit{Varying concentration} --- All of our results so far have focused on nanoclusters in MLG and BLG systems with the same lateral dimensions. However, both the stacking fault energy for poor stacking between the graphene sheets, and the ability of interstitial Ru atoms to weaken the interlayer carbon bonds and shift adjacent graphene layers are naturally extensive properties, i.e., they  will vary with graphene lateral cell dimensions. We have therefore performed additional calculations of the intercalation energy of a single Ru adatom in AA and AB BLG with different graphene supercell sizes. We thus vary both the distance between intercalated Ru atoms and, equivalently, the density of Ru in the graphene sheet. These results are shown in Fig. 5, which plots the effective stacking fault energy in the presence of intercalated single Ru atoms as the difference 
\begin{equation}
    E_{sf} = E_{AA} - E_{AB},
\end{equation}
between the AA-BLG and AB-BLG structures, which is positive when AA stacking is preferred, and negative for AB stacking. While AB stacking is preferred at low concentrations, above a critical value of $\rho \approx 0.005 \rm{\AA}^{-2}$, we find the transition to AA BLG will occur.

This behaviour suggests that Ru atoms will preferentially migrate towards basal (interlayer) dislocations and stacking faults. In addition, it should also result in a stacking fault in ruthenium rich regions, which will return to perfect stacking through the passage of a basal dislocation. The typical energy of such a dislocation is around $E_{b} = 7$  meV/{\AA} \cite{Dai2016}. Taking the Ru-induced stacking fault energy as a function of density to be $E_{sf}(\rho)$, a circular region of intercalated ruthenium atoms will shear the adjacent layers under the condition that 
\begin{equation}
    2 \pi r E_{b} < \pi r^2 E_{sf}(\rho).    
\end{equation}
To illustrate this, in Fig. 5 we also show the critical loop radius above which a stacking shift is initiated. Notably, this loop radius is significantly smaller than the graphite basal dislocation width \cite{Dai2016, McHugh2021}, and so the actual minimum radius must necessarily be higher in order to facilitate the formation of well-defined dislocation cores.

\begin{figure}[htbp]
\includegraphics[width=0.45\textwidth, height=\textheight, keepaspectratio]{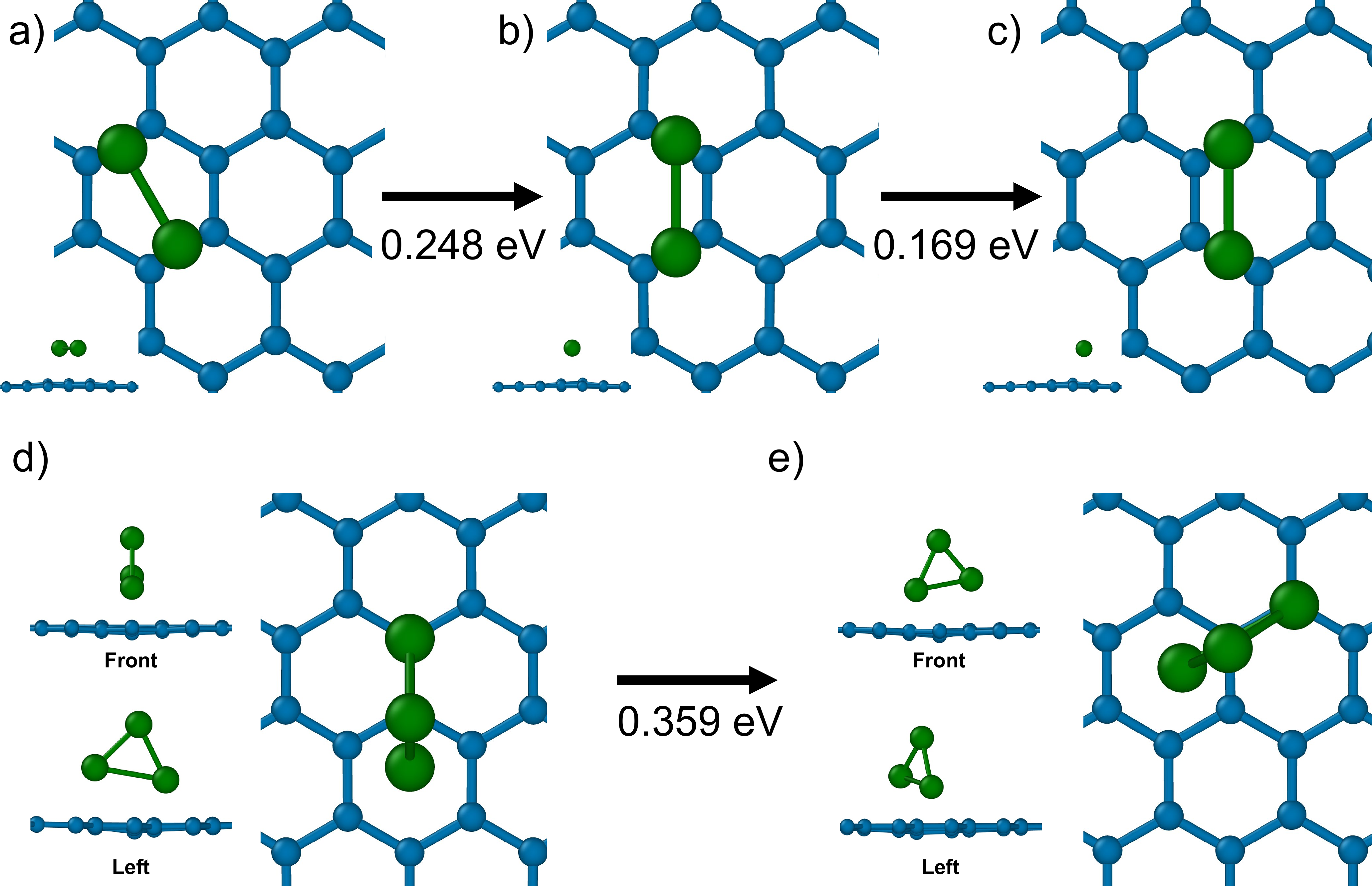}
\caption{NEB calculated diffusion pathways for $n=2,3$ clusters on a graphene sheet. (a) BB1 Initial state (b) this can rotate to the BB2 intermediate state and (c) translate to an adjacent BB2 site in a two-step process. (d) Initial and (e) final states of the $n=3$ nanocluster}
\label{fig7}
\end{figure}

\begin{figure*}[htbp]
\includegraphics[width=1\textwidth, height=\textheight, keepaspectratio, trim = 0cm 4cm 0cm 3cm, clip = true]{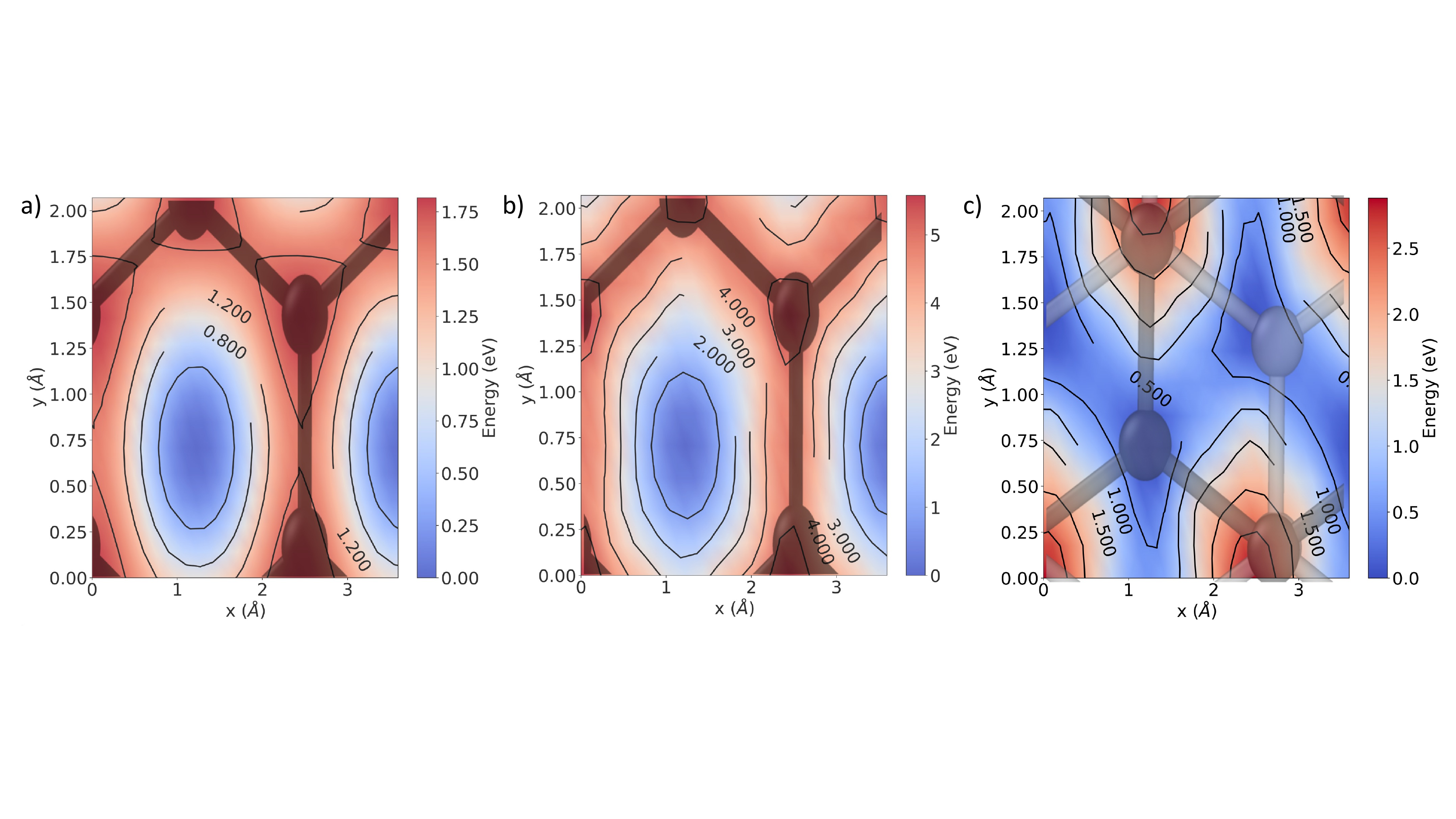}   
\caption{Energy of a single Ru atom as a function of relative displacement (a) adsorbed on graphene, and  (b) intercalated into an AA and (c) an AB bilayer.}
\label{fig8}
\end{figure*}

\begin{figure*}[htbp]
\includegraphics[width=0.85\textwidth, height=\textheight,keepaspectratio]{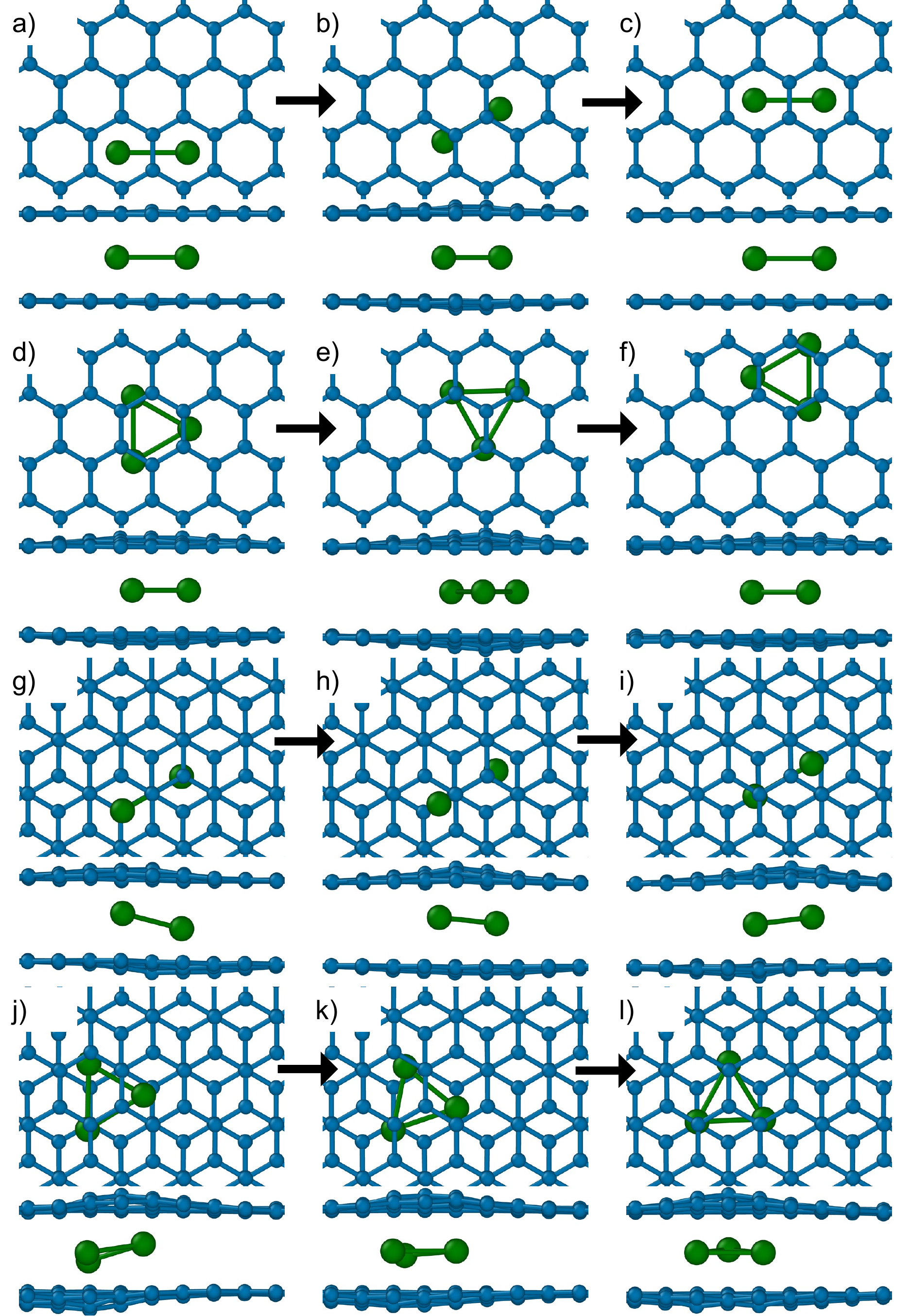}
\caption{Diffusion pathways for Ru clusters in BLG. (a) initial state (b) transition state and (c) final state for $n=2$ Ru cluster in AA BLG. (d) initial state e) transition state and f) final state for $n=3$ Ru cluster in AA BLG. (g) initial state (h) transition state and (i) final state for $n=2$ Ru cluster in AB BLG. (j) initial state (k) transition state and (l) final state for $n=3$ Ru cluster in AB BLG. }
\label{fig9}
\end{figure*}


\subsection{Diffusion}
Thus far we have focused on the structure and energetics of elemental Ru as a function of nanocluster size. However, the penetration of small clusters into the graphite lattice will be determined by dynamical processes which can be understood through calculation of transition states and diffusion barriers. We now examine the diffusivity of the previously discussed clusters, again considering MLG, AA and AB bilayers, and paying particular attention to the effect that interlayer stacking has on diffusive properties. Calculations of  diffusion saddle points have been performed between pairs of the lowest-energy sites discussed previously, and the transition path has been optimised according to the CI-NEB method \cite{neb1, neb2}. The activation energy for the associated paths can then be used to estimate the thermodynamic likelihood of the associated process.

\textit{Single atom} --- We initially consider the diffusion paths taken by $n = 1,2,3$ Ru nanoclusters. Single Ru diffusion paths are shown in Fig. 6, and involve simple movement between adjacent hollow sites, for the monolayer (Fig. 6 (a)) and AA (Fig. 6 (b)) cases, and from a TH-site to an adjacent HT-site for the AB bilayer  (Fig. 6 (c)). There is a notable difference in the diffusion barrier between the three systems. For an Ru atom on single-layer graphene, we find an activation energy of 0.709 eV, which is in very good agreement with our previous results using somewhat different approximations \cite{McHugh2020}.

It is notable that the diffusion barriers within the AA and AB bilayers show opposite tendencies with respect to the barriers for graphene. For the AA stacked bilayer, the barrier is more than doubled with respect to the monolayer to 1.88 eV, while for AB stacking the barrier is significantly decreased to 0.152 eV. These differences are readily understood by a calculation of the intercalation energy of the Ru atom as a function of lateral position above a graphene monolayer, or encapsulated within a bilayer. The result is shown in Fig. 8, where we note a very similar energy surface for the monolayer and AA BLG energies.

In both cases, a single Ru atom pays a significant penalty for sitting either directly above a carbon atom or a C-C bond. The energy at a bond site is marginally lower than above a carbon atom, and it is thus the B-site which serves as the transition state for hopping between H-sites. This energy is effectively doubled for AA BLG due to the second graphene layer. In contrast, barrier heights and energy profiles are significantly reduced for AB BLG, see Fig. 6 (c), where we see that carbon atoms at TT-sites (i.e. one carbon atom above another) are now the highest energy positions, while TH-sites serve as energetic minima. In this case, Ru adatoms can now proceed between adjacent TH-sites without crossing a C-C bond and the activation energy is significantly reduced.

\textit{Dimer \& Trimer} --- Interestingly, the diffusion of $n=2,3$ Ru clusters on a monolayer is generally faster than the $n=1$ case. The relevant processes are shown in Fig. 7. For $n=2$, diffusion is a two-step process: there is an initial rotation from the BB1 to the BB2 configuration, Fig 7 (a) and (b), with $E_{rot} = 0.248$ eV, following which the barrier for translation between adjacent BB2 sites is $E_{trans} = 0.169$ eV. For the trimer, diffusion takes place through a concerted, simultaneous translation and rotation, Fig 7 (d) and (e), during which the top atom has no contact with the graphene sheet. This process also has a barrier of $E_b = 0.359$ eV. As previously discussed for a single Ru atom, it is largely the increased hybridization at the T and B sites which increases Ru energy at these sites, and the lower barriers of the $n=2,3$ clusters can then be understood through the observation that the formation of homo-metallic Ru bonds and decreasing contact with the graphene basal plane, decreases the covalent interaction between Ru atoms and graphene.

In the bilayer case, the most important distinction with the monolayer is that the ability to relax out of plane, and thus to reduce the amount of Ru-C interaction, is greatly inhibited by the surrounding graphene sheets. This has a marked effect on the associated processes. Dimer and trimer diffusion in AA BLG are shown in Fig. 9 (a)-(f), which both proceed via a relatively simple transition state. There is a high degree of contact between the Ru cluster atoms and C atoms at both transition states, yet despite this, the barrier rapidly decreases from $E_b = 0.925$ eV to $E_b = 0.387$ eV. This is highly similar to the behaviour seen in the monolayer, where less covalent interaction with the substrate due to the Ru-Ru cluster bonds aids diffusion.

Conversely, for the diffusion of dimer and trimer clusters in AB BLG, the opposite trend is observed. The transition states in both cases are highly similar to the AA transition states, Fig. 9 (g)-(l), yet the barriers actually increase with increasing cluster size, from $E_b = 0.830$ eV to $E_b = 1.297$ eV. For both clusters, the lowest energy path now requires contact between ruthenium and carbon atoms, which cannot be avoided without paying a significant price for bond stretching and bending, in contrast to single atom diffusion which could entirely avoid these sites.

\textit{Larger clusters} --- CI-NEB transition energy barriers for all cluster sizes on a monolayer and in AA and AB BLG are plotted in Fig. 10 (a) (Additional transition states are shown in the supplementary information). Remarkably, we see almost entirely opposite behaviour for the two stacking configurations, which we can ascribe to the changing degree of hybridization for different cluster sizes. At higher concentration this tendency starts to even out to a larger degree, and with increasingly large ruthenium cluster size, we anticipate that these barriers will start to grow linearly in proportion to the contact area between the Ru and graphene, causing the intercalated cluster to become immobile and begin ripening into larger metallic fragments.

\begin{figure*}[htbp]
\includegraphics[width=0.49\textwidth, height=\textheight, keepaspectratio, trim = 6cm 1cm 6cm 0cm, clip=true]{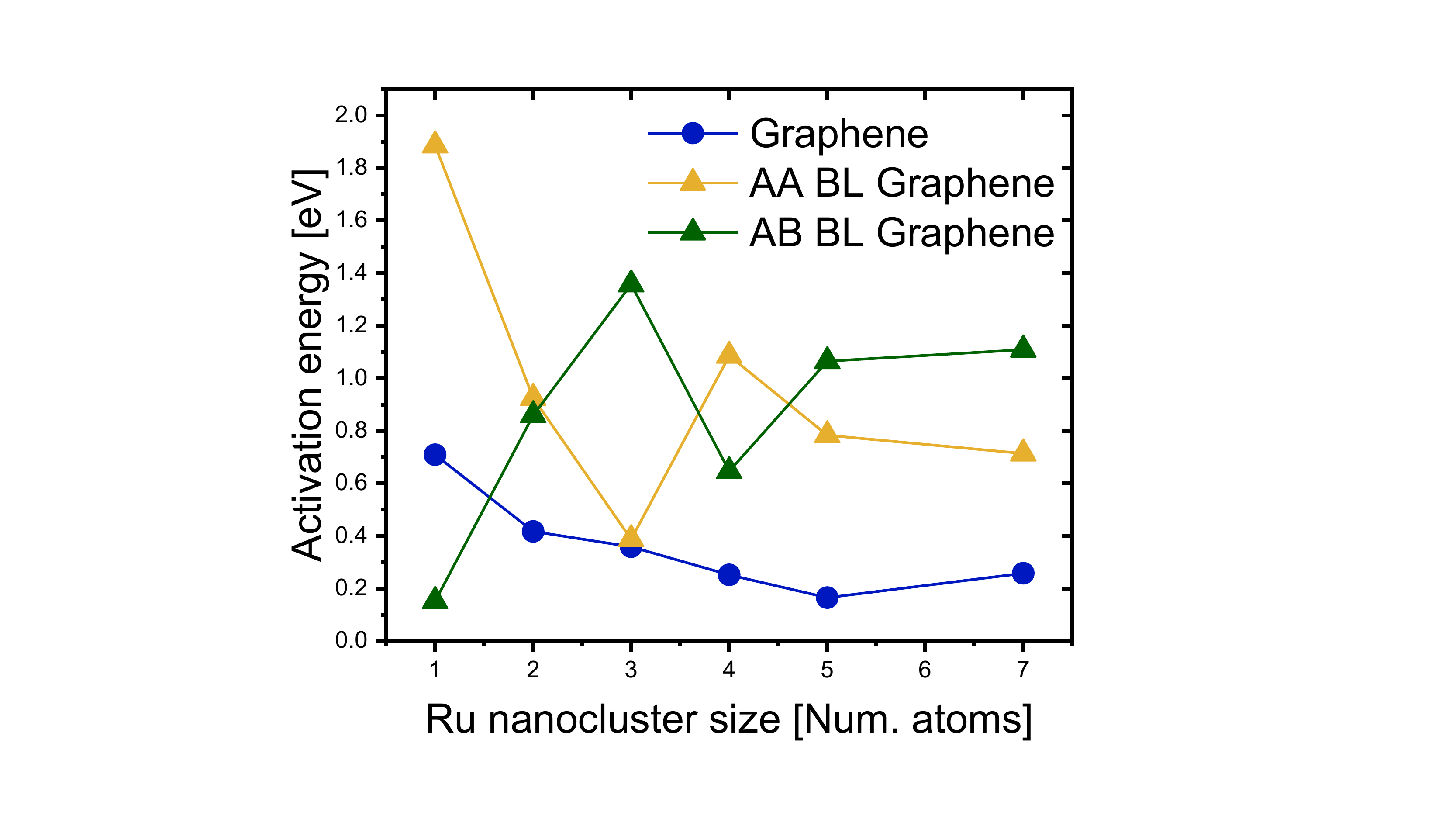}
\includegraphics[width=0.485\textwidth, height=\textheight, keepaspectratio, trim = 6cm 1cm 7cm 0cm, clip=true]{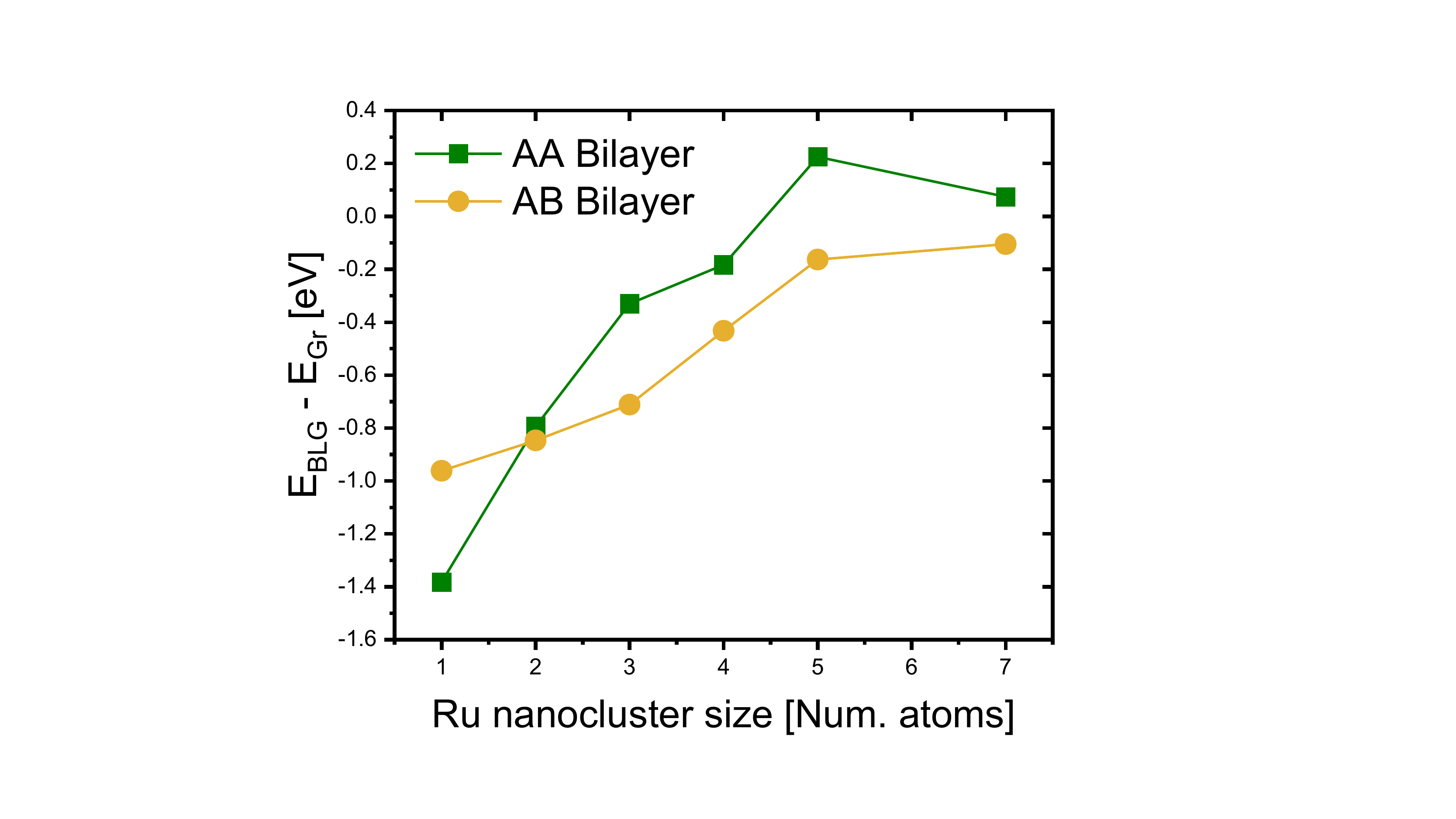}
\caption{Left: CI-NEB activation energy vs. nanocluster size for Ru nanoclusters on MLG, AA and AB BLG. Right: Segregation energy plotted against Ru nanocluster size  for Ru nanoclusters on AA and AB BLGs relative to MLG.}
\label{fig8}
\end{figure*}


\section{Discussion}
In this work, we have performed extensive simulations of the adsorption, intercalation, clustering and diffusion of ruthenium nanoparticles in graphite, which we find has a pronounced impact on structural, energetic and dynamical properties. Most importantly, we observe that interstitial Ru has the capability of shifting the surrounding graphene layers to an AA stacking configuration. This is a particularly interesting effect as it is accompanied by a substantial increase in the transition barrier, which severely impedes diffusion relative to other cluster sizes and stacking configurations. In addition, we find that penetration into the the bulk of graphite is energetically preferably for all Ru nanoclusters sizes, however, this tendency becomes much weaker with increasing cluster size, as shown in Fig. 10 (b), which shows the segregation energy (difference in adsorption and intercalation energies) vs cluster size extracted from our DFT calculations.

We conclude that large clusters are more likely to become trapped at the surface and in porous regions of more disordered graphite grades. Only smaller clusters, which have larger thermodynamic incentive to intercalate, can proceed via defective entry portals into the bulk. Thus, the decreasing thermodynamic preference for intercalation with increasing cluster size necessitates cluster break-up as an active part of the intercalation process. Notably, it has been found recently that the dynamical process underlying the penetration of metal intercalants into the bulk is mediated by a low energy barrier bond-breaking process of the metal dimer at defect sites \cite{Liu2021}. This can effectively filter clustered surface fragments into isolated single atoms, which then proceed to intercalate into graphite. In combination with our observation of higher barriers for single interstitial Ru, and the general preference for these isolated ruthenium atoms for defected regions such as vacancies and basal dislocations, this suggests that the high temperature breakup of clusters into single Ru atoms at entry sites will impede overall diffusion into the bulk and throughout the microstructure.

In general, the modelling of the penetration, diffusion and interaction of ruthenium atoms and clusters with graphite is a challenging one which is vital to many applications and prospective uses of graphite. A fuller understanding of these issues requires larger cells and longer timescales than those which are currently feasibly using DFT. While molecular dynamics would be a promising technique to address this problem, there are currently few interatomic potentials available for many of the most interesting metal dopants, and those which are available typically fail to correctly capture both processes of stacking and of clustering. As this is an important and interesting problem, this would be a fruitful direction for future research.

\textit{Acknowledgements} --- JGM was supported by the UK EPSRC grant EP/R005745/1, Mechanisms of Retention and Transport of Fission Products in Virgin and Irradiated Nuclear Graphite. 
KJ and PM gratefully acknowledge funding from EDF Energy, and the United Kingdom EPSRC grant EP/V050281/1, Modelling long timescale effects of irradiation damage of nuclear graphite.  
The authors gratefully acknowledge the use of the HPCMidlands+ facility, funded by EPSRC grant EP/P020232/1, as part of the HPC Midlands+ consortium.

\nocite{*}
\bibliography{apssamp}


\end{document}